\documentstyle[11pt,newpasp,twoside,epsf]{article}
\def\edcomment#1{\iffalse\marginpar{\raggedright\sl#1\/}\else\relax\fi}
\reversemarginpar

\def\bea{\begin{eqnarray*}}
\def\eea{\end{eqnarray*}}
\def\ba{\begin{eqnarray}}
\def\ea{\end{eqnarray}}

\begin{document}
\title{Multi-wavelength observations of the TeV Blazars Mkn~421,
1ES1959+650, and H1426+428
 with the
HEGRA Cherenkov telescopes and the RXTE X-ray
satellite}
\author{D. Horns for the HEGRA collaboration}
\affil{Max-Planck-Institut f\"ur Kernphysik, Postfach 10\,39\,80, D-69117
Heidelberg, Germany}
\begin{abstract}
 Recent results obtained with the HEGRA system of imaging
Cherenkov telescopes  on the TeV emission of the Blazars Mkn~421 ($z=0.030$), 
1ES1959+650 ($z=0.047$), and  H1426+428 ($z=0.129$) are reported.
For Mkn~421, a close connection of the average flux level and spectral
shape has been observed during the periods of increased activity in 
the years 2000 and 2001. Simultaneously taken data with the RXTE 
X-ray satellite reveal a complex light curve at X-ray and TeV energies.
After a deep exposure of 94~hrs, the object 
1ES1959+650 was detected 
at the significance level of 5.4$\sigma$ with a soft energy spectrum
following a power-law with a photon-index of $3.3 \pm 0.7$. 
During recent observations in May 2002, the source has shown 
increased activity with indications for a   flattening of the energy spectrum. 
  The high energy peaked
Blazar H1426+428 has recently been identified as a source of
TeV photons.  Since the source is fairly distant (z=0.129), absorption of
TeV photons due to pair-production on the optical and near
infrared extragalactic light becomes important and should
leave a signature in the observed TeV energy spectrum.
Notably, the TeV energy spectrum determined with the
HEGRA system of Cherenkov telescopes agrees with the
expectation of a strongly absorbed source spectrum.\end{abstract}

\section{Introduction}
 The surprisingly large number of Active Galactic Nuclei
 discovered by
the EGRET instrument on-board the Compton Gamma Ray observatory has fueled
interest in the search for TeV emission from these sources. Particularly, the 
Blazars (flat radio spectrum quasars and BL Lac objects) 
among the EGRET identified sources have raised the interest of
observers at energies accessible to ground based imaging air Cherenkov
telescopes (IACTs). 
All currently known extragalactic sources of TeV 
$\gamma$-radiation belong to the class of nearby BL Lac objects. 
These sources are 
believed to produce $\gamma$-rays emanating from a population of 
energetic non-thermal particles which move relativistically along the
jet-axis aligned with the line of sight.
 The nature, origin, and initial acceleration mechanism of
these relativistic particles are however not yet identified. \\ The observations
of Blazars at TeV energies have so far helped to constrain the size of 
the emission region and set lower limits to the relativistic 
Doppler factor that one has to assign to the bulk motion of the emitting
region. \\
 In Section \ref{hegra}, We will summarize results on three 
Blazars with different red shift (Mkn~421, 1ES1959+650, H1426+428)
 obtained with the
HEGRA system of IACTs. Initial results of an extended
RXTE campaign to observe Mkn~421 simultaneously at X-ray and 
TeV energies 
will be reported in Section \ref{rxte}
\section{Observations with the HEGRA CT-System}
\label{hegra}
\subsection{Mkn~421 ($z=0.030$)}
\begin{figure}[ht!]
\plotone{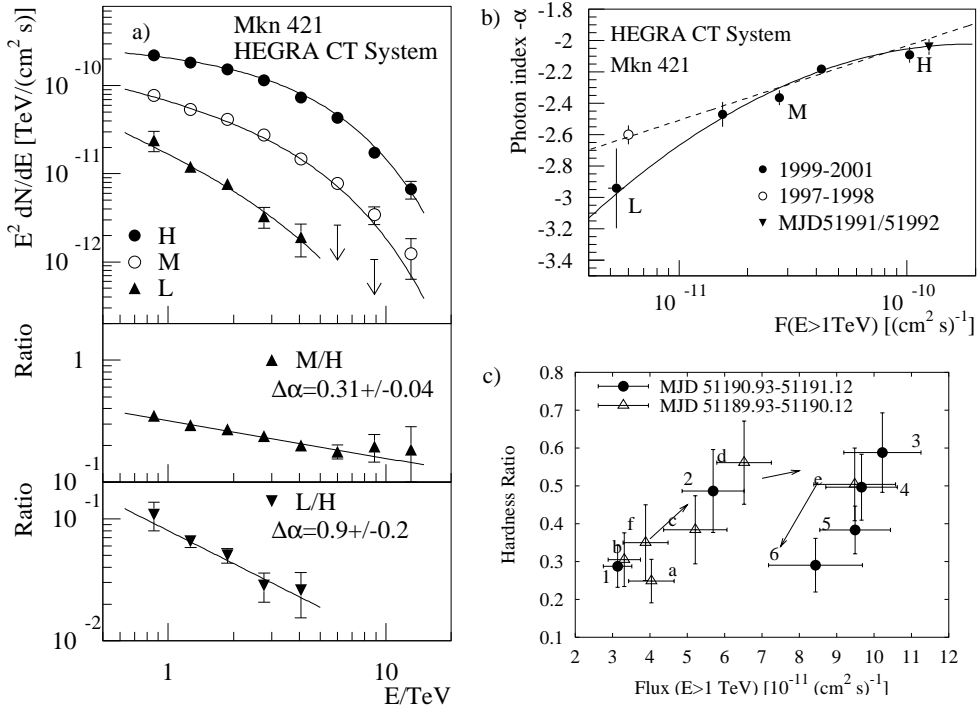}
 \caption{\label{Fig.Mkn421} For different integral flux levels 
observed from Mkn~421, 
spectral fits have been applied to the data (a) and the
ratio of energy spectra have been calculated. 
While keeping the
position of the exponential cut-off constant at 3.6~TeV,  the
photon index shows a correlation with the flux level (right panel).
 For
a comparison, the same fit has been applied to 
archival data of 1997/1998. For two successive nights, variations
of the hardness ratio have been observed, following a clock-wise
pattern in a Hardness-Flux diagram (lower right panel).} 
\end{figure}

 This object has been intensively monitored with the HEGRA IACTs 
since 1997 (Aharonian et al. 1999). Until June
2000, the source has remained at a low flux level.
However, strong activity has been observed during the years 2000 and 2001, where
the source remained at an overall flux level of approximately twice the
flux of the Crab Nebula (Aharonian et al. 2002b).  The 
energy spectra
for different average flux levels have been derived (Fig.~\ref{Fig.Mkn421}a),
indicating at all flux levels a curvature well described by an exponential
cut-off at $3.6\pm0.3$~TeV. 
The observations reveal clear
evidence for a correlation of the spectral shape and the
flux level. The  spectrum becomes harder during higher 
fluxes and softens at lower fluxes. The correlation of photon
index and flux is shown in Fig.~\ref{Fig.Mkn421}b. 
 We have observed diurnal variations of the energy spectrum. 
Plotting the hardness-ratio and integral fluxes in a diagram, we see 
a clock-wise behavior in the hardness-ratio vs. flux diagram (see
Fig~\ref{Fig.Mkn421}c).

\subsection{1ES1959+650 ($z=0.047$)}
\begin{figure}[ht!]
\plotfiddle{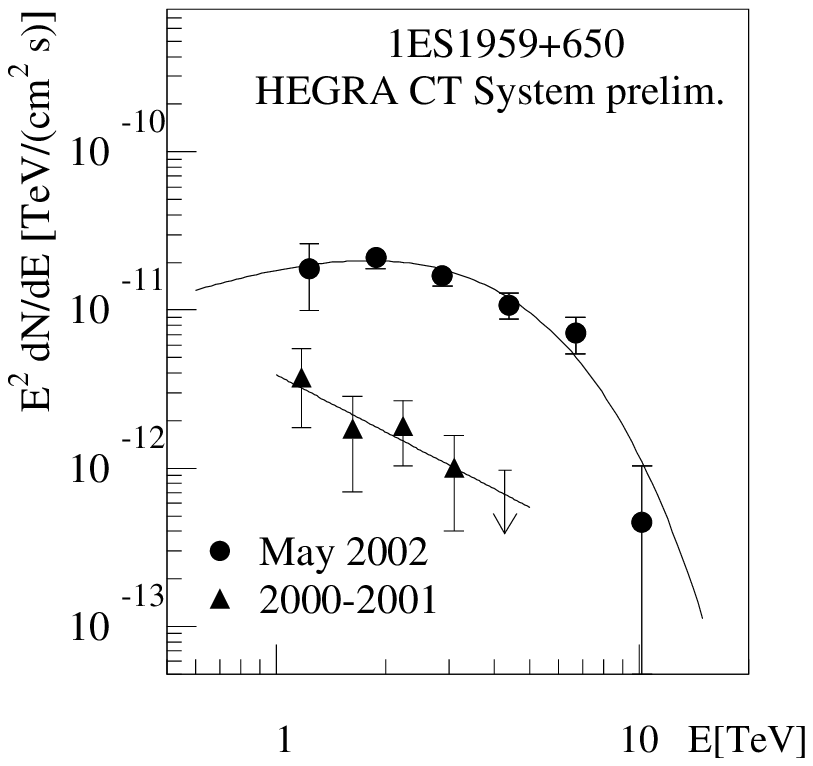}{8cm}{0}{100}{100}{-130}{0}
\vspace*{-0.8cm}
\caption{\label{FIG.1ES1959} For different average flux levels, the
spectrum of 1ES1959+650 shows indications for differences in the spectral shape.}
\end{figure}

A deep observation with HEGRA 
of 94~hrs in 2000 and
 2001 has revealed this object as a weak TeV source, confirming
earlier tentative claims (Nishiyama et al. 1999).
 The integral flux
level averages at 5\% of the flux observed from the Crab-Nebula. 
The significance of the detection is at $5.4\sigma$. 
The energy spectrum extracted from these
data shows a steep source spectrum well described by a power-law with 
photon index $\alpha=3.2\pm0.3$ (Aharonian et al. 2002c).
 Recent observations in May~2002 
showed strong variations of the flux level with peak values being
more than an order of magnitude higher than the previous detection in the
years 2000 and 2001.  The preliminary energy spectrum during the flare 
exhibits pronounced curvature and deviates
significantly from the steep spectrum seen during the quiescent state (see
Fig.~\ref{FIG.1ES1959}). 

\subsection{H1426+428 ($z=0.129$)}
\begin{figure}
\plotfiddle{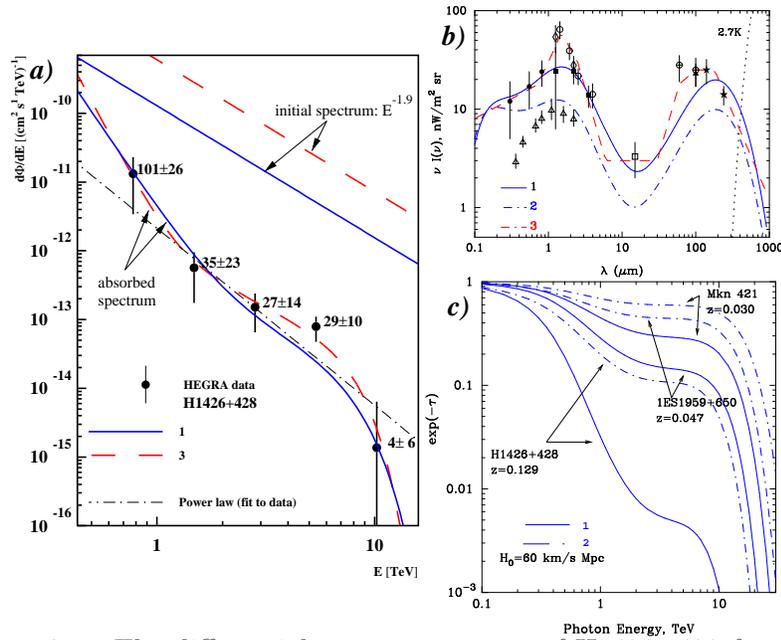}{7.3cm}{0}{75}{75}{-150}{-14}
\caption{\label{Fig.BL1426} The differential energy spectrum
of H1426+428
from the 40~hrs of observations with HEGRA. On the right, 
the SED of the Optical/Infrared extragalactic background light
is shown together with the extinction for TeV photons. The different curves
are explained further in the text.}
\end{figure}

 This object was discovered by HEAO observations and was later 
identified as a BL Lac object 
and confirmed as a steady and bright X-ray source in Einstein 
and EXOSAT observations
(Remillard et al. 1989).
 Furthermore, during Beppo-SAX observations in 1999 (Costamante et al. 2001),
 the X-ray spectrum has been found to be very hard and
to extend up to $\approx$100~keV without indication of a steepening. 
 The source is a member of the class of extreme Synchrotron BL Lacs 
(Costamante et al. 2001) with features similar to those observed at X-rays
from Mkn~501 during the high-state in 1997 (Pian et al. 1998).
The HEGRA system of IACTs has been used to 
observe the source in 1999 and 2000 for a total of 44.4~hrs. An excess with a significance of $5.8\sigma$ from the direction of H1426+428 reveals
this source as an emitter of TeV emission. No strong indication
of variability was found (Aharonian et al. 2002a).\\
  The energy spectrum has been extracted (Fig.~\ref{Fig.BL1426}a). 
Given the red-shift of the object of $z=0.129$ and the expected
diffuse extragalactic background light (EBL) at wavelengths between
1 and 15~$\mu$m (see Fig.~\ref{Fig.BL1426}b),
 considerable absorption of the TeV  photons is to be
expected (see Fig.~\ref{Fig.BL1426}c).
 However, given the uncertainty of the currently available
measurements and estimates of the strength of the EBL, the 
optical depth at 5 TeV is roughly constrained to be $2<\tau<5$. In any case,
the observed spectrum is expected to be modified by absorption features. 
We have adapted a model for the EBL similar to one calculated 
by Primack (2001) to a current compilation of measurements. 
This model of the EBL predicts a steep slope of the SED between 
2 and 15~$\mu$m. The steep slope produces a flattening of the 
optical depth between 2 and 8~TeV which would lead to a \textit{flattening}  of the 
observed spectrum (see Fig.~\ref{Fig.BL1426}c). Indeed, 
such a flattening is consistent with the
observed data points (see again Fig.~\ref{Fig.BL1426}a). Remarkably, the
~5~TeV data point itself shows the highest significance which  is 
consistent with a very hard spectrum in this energy-region. 
The recently published
spectral data at lower energies (Petry et al. 2002, Djannati-Ata\"\i\, et al.
2002) hint at a steep slope with a photon index of $3.5$ below $\approx
2$~TeV. This would be consistent with our low energy data and  would 
support our 
claim about a spectral flattening at higher energies as it is expected
for a heavily obscured TeV-spectrum.

\section{Observations with RXTE}
\begin{figure}
\plotfiddle{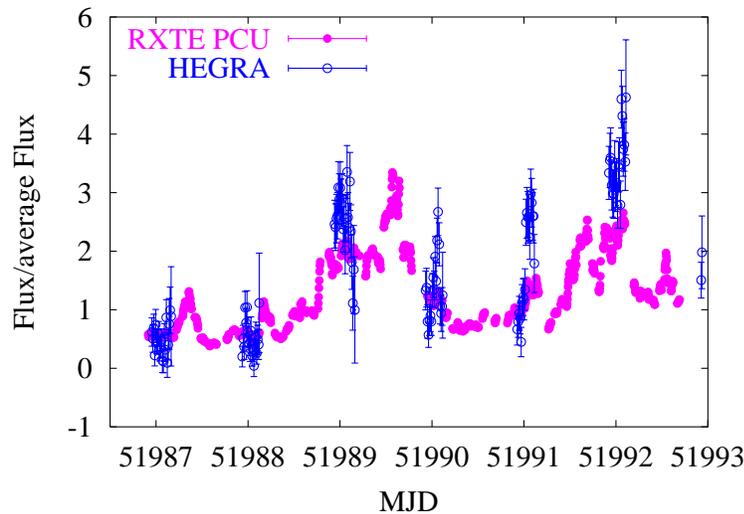}{5.8cm}{0}{80}{80}{-190}{-50}
\caption{The combined TeV and X-ray light-curves in arbitrary units for the 
\label{RXTE}
RXTE campaign on Mkn~421. The flux values have been normalized to the average value for each
energy region.}
\end{figure}

 In March 2001, the RXTE pointed instruments have been used to 
monitor Mkn~421 continuously for one week. During this time, the
flux showed strong flux and spectral variability  at TeV and
X-ray energies. A more in depth analysis of the data are in preparation. Here, we
present the combined X-ray and TeV light-curve of Mkn~421 during the
monitoring (see Fig.~\ref{RXTE}). The flux variations show a complex
pattern at both energy regions. Notably, a tight correspondence
of flaring activities is not evident.
\label{rxte}
\section{Conclusion}
 The HEGRA system of Cherenkov telescopes continues to successfully explore
the wealth of Blazar phenomena at the highest energies.  So far, four
BL Lac objects have been detected by HEGRA and the energy spectra have
been studied in detail.
 Here, we have presented results on three of the four BL Lac objects
(not including Mkn~501)
that have been observed extensively with HEGRA.  For Mkn~421, spectral variations
correlated with the flux level have been observed. The simultaneous
observations at X-ray and TeV energies indicate a complex light curve with
only little correspondence of the flaring behaviour at the two energies.
 The newly discovered source
1ES1959+650 has been detected at the level of $5.4¨\sigma$ during a deep
observation in 2000 and 2001. This source has shown strong and unprecedented
flares in 2002 accompanied by indications for a flattening of the 
observed energy spectrum similar to what has been observed for Mkn~421. 
The most distant TeV source H1426+428 ($z=0.129$) has been detected
and the observed energy spectrum is consistent with absorption features 
caused by pair-production processes with the diffuse extragalactic 
background light between $1-15\,\mu$m.


\end{document}